\documentclass[10pt,aps,pra,twocolumn,showpacs,amsmath,amssymb,footinbib]{revtex4-1}

\usepackage{graphicx}	
\usepackage{dcolumn}	
\usepackage{bm}				
\usepackage{color}		
\usepackage{hyperref}

\begin{document}

\title{Analysis of photon-mediated entanglement between distinguishable matter qubits}

\author{A. M. Dyckovsky${}^{1,2}$}
	\email{aridyckovsky@gmail.com}
\author{S. Olmschenk${}^{1}$}
	\affiliation{${}^{1}$Joint Quantum Institute, University of Maryland Department of Physics and
National Institute of Standards and Technology, College Park, MD 20742, USA \\
	${}^{2}$Loudoun County Public Schools Academy of Science, Sterling, VA 20164, USA}

\date{\today}

\begin{abstract}
We theoretically evaluate establishing remote entanglement between distinguishable matter qubits through interference and detection of two emitted photons. The fidelity of the entanglement operation is analyzed as a function of the temporal and frequency mode-matching between the photons emitted from each quantum memory. With a general analysis, we define limits on the absolute magnitudes of temporal and frequency mode-mismatches in order to maintain entanglement fidelities greater than 99\% with two-photon detection efficiencies greater than 90\%. We apply our analysis to several selected systems of quantum memories. Results indicate that high fidelities may be achieved in each system using current experimental techniques, while maintaining acceptable rates of entanglement. Thus, it might be possible to use two-photon-mediated entanglement operations between distinguishable quantum memories to establish a network for quantum communication and distributed quantum computation.
\end{abstract}

\maketitle

\section{Introduction}
\label{sec:introduction}
Quantum information promises extraordinary advances over classical means of both communication and computation. Quantum communication offers the potential to securely transfer information over long distances \cite{kimble:qinternet}, while quantum computation may enable processing tasks that are intractable using classical methods, such as efficient factorization and quantum simulations of many-body systems~\cite{ladd:qc_expt_review}. The resources required for both of these tasks may be established by photon-mediated long-distance probabilistic entanglement between quantum memories. Quantum memories allow photon-mediated entanglement protocols to circumvent the detrimental exponential scaling associated with direct transmission~\cite{DLCZ2001,quantum-memory-review2010,BDCZ1998}. While remote entanglement has been demonstrated between identical quantum memories~\cite{chou:remote_ensemble, matsukevich:remote_ensemble, moehring:ion-ion}, using distinguishable memories may be unavoidable when relying on fabricated devices such as optical cavities, solid-state qubits, and spectral filters. The entanglement of distinguishable quantum memories may also allow the favorable characteristics of disparate systems to be advantaged. For instance, solid-state quantum memories enable fast qubit operations and may be fabricated using standard methods that include integration of an optical cavity to increase photon collection efficiency and facilitate scaling a quantum network. On the other hand, atomic quantum memories provide long information storage times (seconds or more) \cite{langer:long-lived_qubit, balabas:long_atom_vapor_relax} that may be necessary for quantum information processing. 

Thus far, remote entanglement of hybrid quantum memories has been achieved between a cavity-coupled atom and a Bose-Einstein condensate \cite{rempe:hybrid_atomBEC_2011}. There have also been recent proposals for the remote entanglement of a quantum dot and a trapped ion using single-photon interference \cite{waks-hybrid}, the local entanglement of ultracold degenerate gases with cantilevers \cite{joshi:entangled_gases}, as well as remote entanglement between superconducting and atomic qubits \cite{taylor:hybrid_interface}. With the aim of hybrid entanglement, two-photon interference between single photons from a quantum dot and parametric down-conversion in a nonlinear crystal has been demonstrated \cite{solomon:hybrid_interference}. While each of these techniques has various applications in quantum information, it may be more beneficial to consider passively robust entanglement protocols between remote quantum memories.

We analyze a generic protocol to establish remote entanglement between a pair of distinguishable quantum memories using a two-photon interference scheme. Unlike single-photon entanglement protocols \cite{waks-hybrid}, two-photon interference schemes are not interferometrically sensitive to the optical pathlength \cite{simon:pathlength-sensitivity}, which may permit practical long-distance entanglement. Although a trade-off exists between fidelity and efficiency for distinguishable sources, this two-photon-mediated entanglement protocol has the potential to reach high fidelity while maintaining an acceptable rate of entanglement. Therefore, it might be possible to use remote entanglement operations between distinguishable quantum memories to establish a quantum repeater architecture \cite{BDCZ1998} for practical long-distance applications in quantum communication and quantum computation.

In the following sections we overview the implementation of the remote entanglement operation between two distinguishable quantum memories, and characterize the fidelity of the protocol through a general analysis of the two-photon interference. We then apply this protocol to a few selected systems for the entanglement of accessible pairs of quantum memories: two spectrally filtered atomic qubits, two cavity-coupled solid-state qubits, and a single solid-state qubit and a single atomic qubit. These schemes may be implemented using previously demonstrated experimental techniques, and thus may be accessible by experiments in the near-term.

\section{Proposed Entanglement Protocol}
\label{sec:initial-protocol}
\begin{figure}
	\centering
	\includegraphics[width=1.0\columnwidth,keepaspectratio]{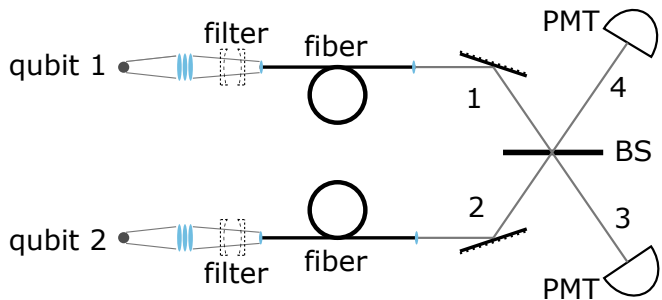}
	\caption{Setup to establish remote hybrid entanglement between qubit 1 corresponding to a photon in spatial mode $m = 1$ and qubit 2 corresponding to a photon in spatial mode $m = 2$.  Photons from each quantum memory are collected through objective lenses, coupled into single-mode fibers, and directed toward the beamsplitter (BS). The photons interfere at the BS, and then photomultiplier tubes (PMTs) detect the photons in spatial modes 3 and 4 in a given window of time. The frequency filters are optional and are not part of the general analysis in Secs.~\ref{sec:initial-protocol} and \ref{sec:photon-interference-analysis}, but are used in analysis of selected systems in Sec.~\ref{sec:results}.}
	\label{fig:entanglement-protocol}
\end{figure}
The envisioned basic setup for remote entanglement of distinguishable quantum memories is illustrated in Fig.~\ref{fig:entanglement-protocol}, which closely resembles proven experimental implementations \cite{moehring:ion-ion,olmschenk:quantum_logic,maunz2009-quantumgate,yuan:atomic_ensembles}.  Here the quantum memories may be separated by an arbitrary distance, limited only by the attenuation length of photons transmitted through the optical fibers. After initialization and excitation of each quantum memory, the spontaneously emitted photons are coupled into the optical fibers and are directed to interfere at a 50:50 nonpolarizing beamsplitter (BS).  The interference and detection of these photons may be used to project the quantum memories into an entangled state.

\begin{figure}
	\centering
	\includegraphics[width=1.0\columnwidth,keepaspectratio]{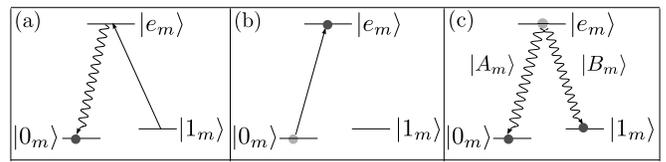}
	\caption{Generic matter-photon entanglement for a single qubit. (a) Optical pumping transfers population into the $\vert 0_{m}\rangle$ ground state for initialization of a qubit $m$. (b) A laser resonant to the $\vert 0_{m}\rangle$ to $\vert e_m \rangle$ transition excites the qubit. (c) The excited state $\vert e_m \rangle$ will decay into a superposition of the ground spin states $\vert 0_{m} \rangle$ and $\vert 1_{m} \rangle$ while spontaneously emitting a photon in a superposition of orthogonal states $\vert A_m\rangle$ and $\vert B_m \rangle$, establishing matter-photon entanglement.}
	\label{fig:total-structure}
\end{figure}

Entanglement can be observed between photon and matter qubits through a sequential method of qubit initialization, excitation, spontaneous emission, and detection. Initialization of each qubit (for instance by optical pumping) prepares the qubit in the $\vert 0_{m}\rangle$ state depicted in Fig.~\ref{fig:total-structure}(a), where $m$ denotes the input spatial mode of a photon entering the BS corresponding to qubit $m$. A laser pulse excites the quantum memory from $\vert 0_{m}\rangle$ to $\vert e_{m} \rangle$ in Fig.~\ref{fig:total-structure}(b). In Fig.~\ref{fig:total-structure}(c), $\vert e_{m} \rangle$, with excited state lifetime $\tau_m$, spontaneously decays into a superposition of qubit states $\vert 0_{m} \rangle$ and $\vert 1_{m}\rangle$, while emitting a photon in a superposition of orthogonal states $\vert A_m \rangle$ and $\vert B_m \rangle$ (e.g. polarizations or frequencies). The matter-photon entanglement between the quantum memory and its emitted photon may be described as $\vert{\psi_m}\rangle=\frac{\vert{0_m A_m}\rangle+\vert{1_m B_m}\rangle}{\sqrt{2}}$. 

After the photons are emitted from each qubit in Fig.~\ref{fig:entanglement-protocol}, a product state $\vert{\Psi}\rangle = \vert{\psi_1}\rangle \vert{\psi_2}\rangle$ exists between the two matter-photon systems. As illustrated in Fig.~\ref{fig:entanglement-protocol}, the photons interfere at the BS, and are then detected in a given window of time by the photomultiplier tubes (PMTs). A coincident detection at the PMTs ideally signals the observation of the post-selected maximally entangled Bell state $\vert{\psi^{-}_{h\nu}}\rangle=\frac{\vert{A_{3}B_{4}}\rangle-\vert{B_{3}A_{4}}\rangle}{\sqrt{2}}$ of the photons. The detection of photons in the $\vert{\psi^{-}_{h\nu}}\rangle$ state heralds the probabilistic entanglement of the matter qubits. The state of the two quantum memories is then $\vert{\psi^{-}_{qub}}\rangle=\frac{\vert{1_{1} 0_{2}}\rangle-\vert{0_{1} 1_{2}}\rangle}{\sqrt{2}}$, which represents remote spin-spin entanglement between qubits 1 and 2 \cite{simon:pathlength-sensitivity,zukowski:bell,braunstein:bell}.

After creating the $\vert{\psi^{-}_{qub}}\rangle$ entangled state, the states of the entangled qubits may be read out. The process to detect each qubit state is dependent upon the structure of each specific quantum memory and will be detailed for each selected system in Sec~\ref{sec:results}.

\section{Photon Interference Analysis}
\label{sec:photon-interference-analysis}

The fidelity of the above protocol depends critically on the quality of the interference between photons, as evaluated below. While the general interference analysis parallels to some extent the work in Ref.~\cite{legero2003}, we have extended the analysis to characterize entanglement fidelity in the general case of two qubits distinguishable by both their frequency and temporal emission profiles. Initially, a single, spontaneously emitted photon can be expressed by either a temporal or frequency mode profile \cite{legero2003}. The overlap of these profiles in the temporal and frequency modes determines the quality of two-photon interference, and can be evaluated as the probability of coincident detection at the PMTs depicted in Fig.~\ref{fig:entanglement-protocol}. We use this probability of coincident detection to evaluate the fidelity of the matter qubit entanglement following joint detection of the interfered photons, as well as the detection efficiency.

Embedded in this analysis is the consideration of potential errors associated with the interference of photons from distinguishable quantum memories, including temporal and frequency offsets. Temporal offsets can be either a difference in arrival time of each photon at the BS or a difference in excited state lifetimes, which determine the temporal profiles of each photon. We incorporate differences in excited state lifetimes in both the present and later sections. Differences in arrival times can be optimized by simply changing the photon pathlength to the BS, which becomes important for analysis in Secs.~\ref{sec:barium} and \ref{sec:hybrid}. A frequency offset is a difference in the mean values of the photon frequency mode profiles. Our analysis below demonstrates that a frequency offset may be tolerated to some extent and may be reduced by using either cavity-coupling or spectral filtering.

To begin, we consider a single emitted photon from a quantum memory that may be described in the temporal mode by the electric field operator  
\begin{equation}
\hat{E}^{+}_{m}(t) = \frac{e^{-\frac{t}{2\tau_m}-i\omega_m t}\Theta(t)}{\sqrt{\tau_m}} \hat{a}_m,
\label{eq:general-operator}
\end{equation}
where $\omega_m$ is the mean frequency of a photon emitted from quantum memory $m$, $\Theta(t)$ is the Heaviside function, and $\hat{a}_m$ is the usual photon annihilation operator. 

A temporal mode profile can be expressed for spontaneously emitted single photons as
\begin{eqnarray}
P_m(t) &=& \langle\psi_m\vert \hat{E}^{-}_{m}(t) \hat{E}^{+}_{m}(t) \vert\psi_m\rangle\notag\label{eq:nvc-temp-profile}
\\
&=& \frac{e^{-\frac{t}{\tau_m}}\Theta(t)}{\tau_m},
\end{eqnarray}
which is the time-dependent probability of spontaneous emission from quantum memory $m$ where $\vert\psi_m\rangle = \hat{a}_m^\dagger\vert 0 \rangle$.

The interference of two photons is evaluated in terms of a joint detection probability (JDP), which describes the likelihood of coincidentally detecting photons after interference at a BS. Here we consider two cases of the JDP, which are then compared to evaluate the quality of the photon interference. First, the case that a time-dependent JDP considers noninterfering photons (e.g. orthogonally polarized photons) reduces to \cite{legero2003,olmschenk:quantum_logic,hong-ou-mandel}:
\begin{widetext}
\begin{equation}
\label{eq:early-dist-jdp}
P_{j,D}(t,t_d) = \langle\psi_{1a,2b}\vert \hat{E}^{-}_{3a}(t) \hat{E}^{-}_{4b}(t+t_d) \hat{E}^{+}_{4b}(t+t_d)\hat{E}^{+}_{3a}(t) \vert\psi_{1a,2b}\rangle + \langle\psi_{1a,2b}\vert \hat{E}^{-}_{3b}(t) \hat{E}^{-}_{4a}(t+t_d) \hat{E}^{+}_{4a}(t+t_d)\hat{E}^{+}_{3b}(t) \vert\psi_{1a,2b}\rangle,
\end{equation}
\end{widetext}
where $a$ and $b$ denote orthogonal states with respect to each other, $\vert\psi_{1a,2b}\rangle = \hat{a}^{\dagger}_1 \hat{b}^{\dagger}_2 \vert 0\rangle$ is a state containing two completely distinguishable photons (each in a different spatial mode $m=1,2$), and time delay $t_d$ is the interval of time between successive detection events at the PMTs depicted in Fig.~\ref{fig:entanglement-protocol}. The electric field operators in Eq.~\ref{eq:early-dist-jdp}, written in terms of the output spatial modes $m=3$ and $m=4$ of the BS depicted in Fig.~\ref{fig:entanglement-protocol} may be written in terms of the input modes as \cite{legero2003}
\begin{equation}
\hat{E}^{\pm}_{3}(t) = \frac{\hat{E}^{\pm}_{1}(t)-\hat{E}^{\pm}_{2}(t)}{\sqrt{2}}, \label{eq:elec-operator-relations}
\qquad
\hat{E}^{\pm}_{4}(t) = \frac{\hat{E}^{\pm}_{1}(t)+\hat{E}^{\pm}_{2}(t)}{\sqrt{2}}.
\end{equation}

Using the electric field operator defined by Eq.~\ref{eq:general-operator}, then Eq.~\ref{eq:early-dist-jdp} can be written as
\begin{widetext}
\begin{eqnarray}
\label{eq:mid-dist-jdp}
P_{j,D}(t,t_d) &=& \frac{1}{2}\langle\psi_{1a,2b}\vert \hat{E}^{-}_{1a}(t) \hat{E}^{-}_{2b}(t+t_d) \hat{E}^{+}_{2b}(t+t_d)\hat{E}^{+}_{1a}(t) \vert\psi_{1a,2b}\rangle + \frac{1}{2}\langle\psi_{1a,2b}\vert \hat{E}^{-}_{2b}(t) \hat{E}^{-}_{1a}(t+t_d) \hat{E}^{+}_{1a}(t+t_d)\hat{E}^{+}_{2b}(t) \vert\psi_{1a,2b}\rangle \notag
\\
&=& \frac{e^{-\frac{t(\tau_1+\tau_2)}{\tau_1\tau_2}}\left(e^{-\frac{t_d}{\tau_1}} + e^{-\frac{t_d}{\tau_2}}\right)\Theta(t)\Theta(t+t_d)}{2\tau_1\tau_2}.
\end{eqnarray}
\end{widetext}
Integrating this over all time $t$ yields the total JDP for the noninterfering case:
\begin{eqnarray}
\label{end-dist-jdp}
P_{J,D}(t_d) &=& \int_{-\infty}^{\infty}P_{j,D}(t,t_d)\,dt\notag
\\
&=& \frac{e^{-\frac{|t_d|}{\tau_1}}}{2(\tau_1+\tau_2)}+\frac{e^{-\frac{|t_d|}{\tau_2}}}{2(\tau_1+\tau_2)},
\end{eqnarray}
which demonstrates that interference does not occur because $P_{J,D}(t_d)\neq0$ for $t_d\in\left(-\infty,\infty\right)$ with no frequency dependence. Because no interference occurs, each photon is transmitted or reflected at the BS with random 50\% probability.

Second, the case that a time-dependent JDP considers interfering photons (e.g. identically polarized photons emitted from the two quantum memories) reduces to \cite{legero2003,olmschenk:quantum_logic,hong-ou-mandel}:
\begin{equation}
P_{j,R}(t,t_d)=\langle\psi_{1,2}\vert \hat{E}^{-}_{3}(t) \hat{E}^{-}_{4}(t+t_d) \hat{E}^{+}_{4}(t+t_d)\hat{E}^{+}_{3}(t) \vert\psi_{1,2}\rangle,
\label{early1-real-jdp}
\end{equation}
where $\vert\psi_{1,2}\rangle = \hat{a}^{\dagger}_1 \hat{a}^{\dagger}_2 \vert 0\rangle$. The electric field operators can be related back to the spatial modes from which the photons were emitted using Eq.~\ref{eq:elec-operator-relations}. With the electric field operators expressed in spatial modes 1 and 2,
\begin{widetext}
\begin{eqnarray}
\label{early2-real-jdp}
P_{j,R}(t,t_d,\Delta\omega) & = & \frac{1}{4}\langle\psi_{1,2}\vert \hat{E}^{-}_{1}(t) \hat{E}^{-}_{2}(t+t_d) \hat{E}^{+}_{2}(t+t_d)\hat{E}^{+}_{1}(t) \vert\psi_{1,2}\rangle + \frac{1}{4} \langle\psi_{1,2}\vert \hat{E}^{-}_{2}(t) \hat{E}^{-}_{1}(t+t_d) \hat{E}^{+}_{1}(t+t_d)\hat{E}^{+}_{2}(t) \vert\psi_{1,2}\rangle\notag \\
														&& - \frac{1}{4}\langle\psi_{1,2}\vert \hat{E}^{-}_{1}(t) \hat{E}^{-}_{2}(t+t_d) \hat{E}^{+}_{1}(t+t_d)\hat{E}^{+}_{2}(t) \vert\psi_{1,2}\rangle - \frac{1}{4}\langle\psi_{1,2}\vert \hat{E}^{-}_{2}(t) \hat{E}^{-}_{1}(t+t_d) \hat{E}^{+}_{2}(t+t_d)\hat{E}^{+}_{1}(t) \vert\psi_{1,2}\rangle\quad\notag \\
														& = & - \frac{e^{-\frac{2t(\tau_1+\tau_2)}{2\tau_1\tau_2}} e^{-\frac{t_d\left(\tau_2+\tau_1\left(1+2i\tau_2\Delta\omega\right)\right)}{2\tau_1\tau_2}}\Theta(t)\Theta(t+t_d)}{2\tau_1\tau_2} + \frac{P_{j,D}(t,t_d)}{2},
\end{eqnarray}
\end{widetext}
where $\Delta\omega = \omega_1-\omega_2$ is the frequency mode mismatch between photons emitted from each quantum memory. The total interfering JDP considering identically polarized photons may be represented as
\begin{widetext}
\begin{eqnarray}
\label{eq:real-jdp}
P_{J,R}(t_d,\Delta\omega) &=& \int_{-\infty}^{\infty}P_{j,R}(t,t_d,\Delta\omega)\,dt\notag
\\
&=& -\frac{1}{2\tau_1\tau_2}\int_{-\infty}^{\infty}e^{-\frac{2t(\tau_1+\tau_2)}{2\tau_1\tau_2}}e^{-\frac{t_d\left(\tau_2+\tau_1\left(1+2i\tau_2\Delta\omega\right)\right)}{2\tau_1\tau_2}}\Theta(t)\Theta(t+t_d) \,dt+ \frac{P_{J,D}(t_d)}{2}\notag
\\
&=& -\frac{ \cos(t_d \Delta\omega)e^{{-\frac{|t_d|(\tau_1+\tau_2)}{2\tau_1\tau_2}}}}{2(\tau_1+\tau_2)} + \frac{P_{J,D}(t_d)}{2},
\end{eqnarray}
\end{widetext}
which demonstrates that interference will always occur when $t_d$ is zero, and interference is still strong at all other values of $t_d$ with near-perfect frequency mode-matching ($\vert\Delta\omega\vert \ll \frac{2}{\tau_1+\tau_2}$).

While we evaluate two-photon interference based on known photon states (parallel or orthogonal), the entanglement protocol depends upon the interference of photons each defined by a superposition state representing matter-photon entanglement as described in Sec.~\ref{sec:initial-protocol}. Thus, the fidelity of the entanglement, or the overlap between the target quantum state and measured quantum state, is determined by the quality of the two-photon interference, since ideally a coincident detection only occurs when the photons are in the $\vert{\psi^{-}_{h\nu}}\rangle$ state. The detection of $\vert{\psi^{-}_{h\nu}}\rangle$ would then swap the entanglement of the photons to the quantum memories. The characteristics of this event are analyzed in terms of fidelity and detection efficiency below.

Fidelity may be derived from maximum and minimum correlated intensities ($I_{max}$ and $I_{min}$, respectively) that accumulate in a window of time $W$ for which the PMTs in Fig.~\ref{fig:entanglement-protocol} are open for detection. The correlated intensities are combined to measure the interferometer visibility
\begin{equation}
\label{eq:visibility}
V(W,\Delta\omega)=\frac{I_{max}-I_{min}}{I_{max}+I_{min}},
\end{equation}
which characterizes the overall quality of the two-photon interference at the BS \cite{olmschenk:quantum_logic}. First, the maximum correlated intensity $I_{max}$ corresponds to the case of noninterfering photons, and is defined as
\begin{eqnarray}
\label{eq:max_intensity}
I_{max} &=& \int_{-W/2}^{W/2}P_{J,D}(t_d)\,dt_d\notag
\\
&=& \frac{1}{\tau_1+\tau_2}\int_{0}^{W/2}\left(e^{-\frac{t_d}{\tau_1}}+e^{-\frac{t_d}{\tau_2}}\right)\,dt_d\notag
\\
&=& \frac{\tau_1\left(1-e^{-\frac{W}{2\tau_1}}\right)+\tau_2\left(1-e^{-\frac{W}{2\tau_2}}\right)}{\tau_1+\tau_2}.
\end{eqnarray}
Second, the minimum correlated intensity $I_{min}$ corresponds to the case of interfering photons, and is defined as
\begin{widetext}
\begin{eqnarray}
\label{eq:min_intensity}
I_{min} & = & \int_{-W/2}^{W/2} P_{J,R}(t_d,\Delta\omega)\,dt_d\notag
\\
				& = & -\frac{1}{\tau_1+\tau_2}\int_{0}^{W/2}\cos{(t_d \Delta\omega)}e^{-\frac{t_d(\tau_1+\tau_2)}{2\tau_1\tau_2}}\,dt_d + \frac{I_{max}}{2}\notag
\\
				& = &\frac{2\tau_1\tau_2 e^{-\frac{W(\tau_1+\tau_2)}{4\tau_1\tau_2}}\left[2\tau_1\tau_2\Delta\omega\sin{\left(\frac{W\Delta\omega}{2}\right)}+(\tau_1+\tau_2)\left(e^{\frac{W(\tau_1+\tau_2)}{4\tau_1\tau_2}}-\cos{\left(\frac{W\Delta\omega}{2}\right)}\right)\right]}{(\tau_1+\tau_2)\left(\tau_1^2\left(4\Delta\omega^2\tau_2^2+1\right)+2\tau_1\tau_2+\tau_2^2\right)} + \frac{I_{max}}{2}.
\end{eqnarray}
\end{widetext}

The fidelity may be represented in terms of visibility as \cite{olmschenk:quantum_logic}
\begin{equation}
\label{eq:fidelity}
F(W,\Delta\omega) = \frac{1}{2-\left(V(W,\Delta\omega)\right)^2},
\end{equation}
where we assume negligible frequency differences between the two relevant transitions within each quantum memory.

The fidelity for any probabilistic entanglement protocol is irrelevant without also evaluating efficiency. Efficiency describes the success probability of the entanglement protocol, and therefore must be high enough for practical quantum communication and quantum computation. The efficiency of two-photon detection 
\begin{eqnarray}
\label{eq:detection}
\eta(W) &=& \int_0^W P_1(t) \,dt \times \int_0^W P_2(t) \,dt\notag
\\
&=& \left(1-e^{-\frac{W}{\tau_1}}\right)\left(1-e^{-\frac{W}{\tau_2}}\right)
\end{eqnarray}
defines the probability of detection within a finite $W$. While this generic detection efficiency assumes neither technical nor experimental losses, the analysis of selected systems below will consider both a net efficiency and an estimated rate of entanglement, which include probabilities of emission, transmission, and detection.

\begin{figure}
	\centering
	\includegraphics[width=1.0\columnwidth,keepaspectratio]{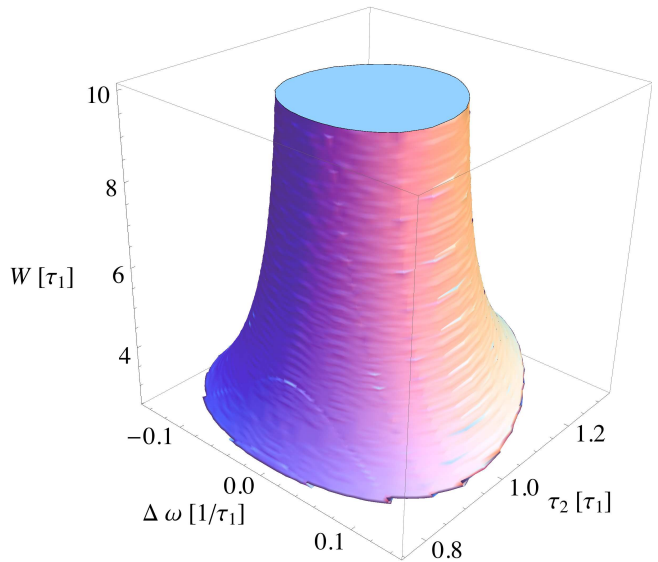}
	\caption{Region representing combinations of $\Delta\omega$, $\tau_2$, and $W$, in units of $\tau_1$, that permit $F(W,\Delta\omega)>99$\% and $\eta(W)>90$\% for the entanglement of two distinguishable qubits. It is apparent that even with substantial photon mode-mismatches high fidelities and detection efficiencies are possible.}
	\label{fig:general_fidelity}
\end{figure}

Both a temporal mode mismatch ($\tau_1\neq \tau_2$) and a frequency mode mismatch ($\Delta\omega \neq 0$) will degrade two-photon interference and thus contribute to entanglement fidelity loss. Although this loss in fidelity can be compensated for by reducing the detection window $W$, this is only possible at the expense of reduced efficiency. In Fig.~\ref{fig:general_fidelity} we show a three-dimensional solid region composed of coordinates $\left(\Delta\omega,\tau_2,W\right)$ in units of $\tau_1$ in which fidelities $>99$\% and two-photon detection efficiencies $>90$\% are possible. As is clearly illustrated, in principle, high fidelities are possible even when using emission sources that differ significantly in both frequency and time.

\section{Protocol and Analysis for Selected Systems}
\label{sec:results}

In Sec.~\ref{sec:photon-interference-analysis} we analyzed the general results and limitations of a two-photon-mediated entanglement protocol when offsets are present in the temporal and frequency modes. That analysis may be applied to the entanglement of several different specific systems of quantum memories that are nominally or experimentally distinguishable, or both. The following subsections describe entanglement protocols for three different selected systems of distinguishable quantum memories: two atomic qubits, two solid-state qubits, and hybrid qubits. Extensions from the general protocol in each scenario are currently feasible with previously demonstrated experimental techniques, showing the current applicability of our analysis in various settings.

\subsection{Two Atomic Qubits}
\label{sec:barium}

\begin{figure*}
	\centering
	\includegraphics[width=2.0\columnwidth,keepaspectratio]{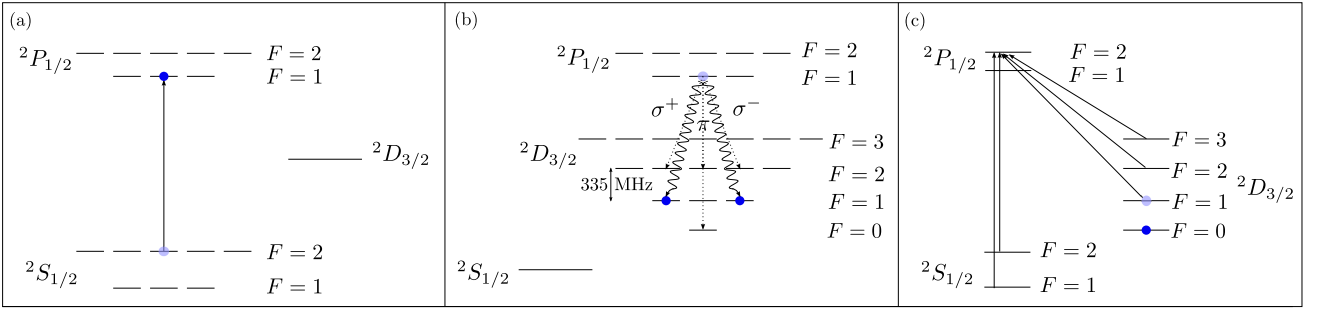}
	\caption{(a) A 493.5 nm laser excites $^{137}$Ba$^{+}$ from the $^{2}S_{1/2}$ $\vert F=2,m_F=0\rangle$ state into the $^{2}P_{1/2}$ $\vert F=1,m_F=0 \rangle$ excited state. (b) With natural lifetime $\tau_{ion}=8$ ns, $^{2}P_{1/2}$ can decay into the $^{2}D_{3/2}$ metastable state by spontaneously emitting a photon at a wavelength of $\lambda_m= 650$ nm. The transition from $^{2}P_{1/2}$ $\vert F=1,m_F=0 \rangle$ to $^{2}D_{3/2}$ $\vert F=1, m_F=\pm 1\rangle$ can establish spin-polarization entanglement between a spontaneously emitted photon and $^{137}$Ba$^{+}$ \cite{villemoes-ba-hyperfine}. The filter \cite{lan2007-filter,nielsen-thesis-filter} from Fig.~\ref{fig:entanglement-protocol} is used to observe transitions only to the $F = 1$ hyperfine level of $^{2}D_{3/2}$. In addition, photons with $\pi$-polarization are excluded by observing along a quantization axis defined by an external magnetic field \cite{blinov-atom-photon}. (c) Read-out of the $^{137}$Ba$^{+}$ qubit spin state. Microwave radiation transfers population from $^{2}D_{3/2}$ $\vert F=1, m_F= +1\rangle$ to $^{2}D_{3/2}$ $\vert F=0, m_F=0\rangle$. Resonant light between $^{2}D_{3/2}$ and $^{2}P_{1/2}$, along with resonant light between $^{2}S_{1/2}$ and $^{2}P_{1/2}$, then produces fluorescence only if the atom is projected into the state $^{2}D_{3/2}\vert F=1, m_F= -1\rangle$.}
	\label{fig:barium-entanglement}
\end{figure*}

The general techniques in Sec.~\ref{sec:initial-protocol} may be used to analyze the entanglement between nominally identical remote atomic qubits, which become distinguishable with experimental fabrication of spectral modifiers. These spectral modifiers, specifically frequency filters, can be necessary to inhibit specific transitions when multiple hyperfine manifold decay channels are present \footnote{It is possible to use a photonic frequency qubit (with $\pi$-polarization) without the requirement of spectral filters, but it may prove beneficial to use $\sigma^{\pm}$ transitions in order to take advantage of currently pursued collection techniques like parabolic mirror geometries \cite{lindlein2007}, which also mitigate errors due to polarization mixing often encountered when utilizing high NA collection optics \cite{luo:scalable_network}.}. The advantage is that transitions to states with such decay channels often provide longer-wavelength emissions, maximizing the attenuation length in optical fiber. Specifically, here we consider each quantum memory (qubits 1 and 2 in Fig.~\ref{fig:entanglement-protocol}) to be a $^{137}$Ba$^{+}$ atom confined in a linear radiofrequency (rf) Paul trap \cite{Paul1990-review}. A spectral frequency filter is included along the path of photons emitted from each trapped $^{137}$Ba$^{+}$ atom to inhibit the observation of unwanted frequencies \footnote{An additional error mechanism that could potentially adversely alter the spectrum of $^{137}$Ba$^{+}$ emission is micromotion caused by the rf trapping fields. However, we consider optimal trapping conditions where proper nulling of static offset fields minimizes micromotion. With standard low temperatures $\approx$ 1 mK the absolute fluctuation in emission frequency due to the secular motion is extremely small \cite{sherman-thesis}. Therefore, these frequency fluctuations of single emitted photons from $^{137}$Ba$^{+}$ should have negligible effects and are omitted from the analysis.}. We have also outlined in Fig.~\ref{fig:barium-entanglement} the processes of excitation, emission, and detection for establishing entanglement after initialization by optical pumping into the $^{2}S_{1/2} \vert F=2, m_F=0 \rangle$ state. The requirements and effects of the above mentioned spectral modifiers are detailed below.

In this protocol, spectral filtering \cite{lan2007-filter,nielsen-thesis-filter} of photons emitted from $^{137}$Ba$^{+}$ is used to observe only the transition from $^2P_{1/2} \vert F=1,m_F=0 \rangle$ to $^2D_{3/2} \vert F=1,m_F=\pm1 \rangle$, suppressing the observation of photons from transitions to $^2D_{3/2} \vert F=2,m_F=\pm1 \rangle$. This can be done with an external Fabry-Perot cavity resonant at the desired transition frequency. The $P$-$D$ transition was chosen as it provides a wavelength more practical than the $P$-$S$ transition for transmission over long-distances. This same technique could also be employed for other atomic ions such as Sr$^+$ or Ca$^+$.

In order to analyze the photon interference in the presence of the two relevant transitions in $^{137}$Ba$^{+}$, we define the electric field operator
\begin{eqnarray}
\label{eq:barium_firstoperator}
\hat{E}^{+}_{m}(t)&=&\frac{e^{-\frac{t}{2\tau_m}-i\omega_m t}\Theta(t)}{\sqrt{\tau_m}}\notag
\\
&\times& \left[\sqrt{L_m}\hat{a}_m+\sqrt{1-L_m}e^{-i\omega_{hfs}t}\hat{b}_m\right],
\end{eqnarray}
where $L_m$ is the natural transition probability to observe the $^2P_{1/2} \vert F=1,m_F=0 \rangle$ to $^2D_{3/2} \vert F=1,m_F=\pm1 \rangle$ transition, $\omega_{hfs}$ is the hyperfine frequency splitting between $^2D_{3/2} \vert F=1\rangle$ and $^2D_{3/2} \vert F=2\rangle$, and operators $\hat{a}_m$ and $\hat{b}_m$ are used to represent the different (distinguishable) transitions.

\begin{figure}
	\centering
	\includegraphics[width=1.0\columnwidth,keepaspectratio]{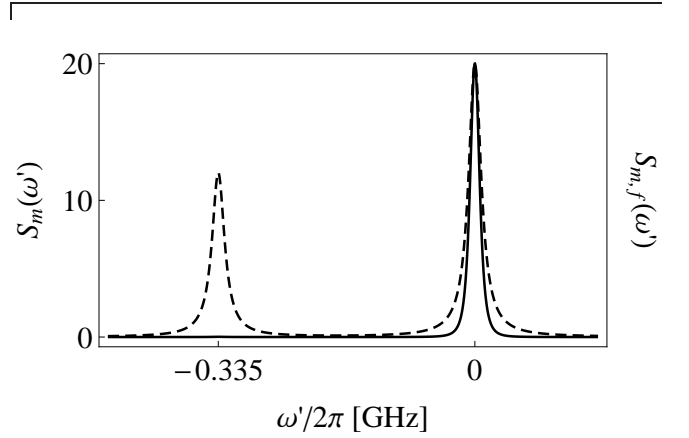}
	\caption{The $\sigma$-polarized emission spectrum of $^{137}$Ba$^{+}$ for the $^2P_{1/2} \vert F=1\rangle$ to $^2D_{3/2}$ transition before (dashed line) and after (solid line) frequency filtering. The transition probabilities are 3/8 to $^2D_{3/2} \vert F=2,m_F=\pm1 \rangle$ and 5/8 to $^2D_{3/2} \vert F=1,m_F=\pm1 \rangle$ \cite{sherman-thesis,metcalf1999}. Here the desired transition to $^2D_{3/2} \vert F=1,m_F=\pm1 \rangle$ is centered at $\frac{\omega'}{2\pi}=\nu'=\nu-\nu_m=0$, the other transition to $^2D_{3/2} \vert F=2,m_F=\pm1 \rangle$ is centered at $\nu_{hfs} = -0.335$ GHz \cite{sherman-thesis}, and $\kappa_m=0.05$ GHz.}
	\label{fig:frequency-filtering}
\end{figure}

To obtain the frequency mode profile of an emitted photon from $^{137}$Ba$^{+}$, we first take the Fourier transform of the electric field operator from Eq.~\ref{eq:barium_firstoperator}
\begin{eqnarray}
\hat{s}_m(\omega') &=& \frac{1}{\sqrt{2\pi}}\int_{-\infty}^{\infty}\hat{E}^{+}_{m}(t)e^{i\omega t}\,dt\label{eq:fourier-frequency-ion}
\\
&=& \sqrt{\frac{2\tau_m}{\pi}}\left[\frac{\sqrt{L_m}}{1-2 i \tau_m \omega'}\hat{a}_m\right.\notag
\\
&&+ \left.\frac{\sqrt{1-L_m}}{1-2 i \tau_m (\omega'-\omega_{hfs})}\hat{b}_m\right]\notag,
\end{eqnarray}
where $\omega' = \omega-\omega_m$. Then the normalized unfiltered frequency mode profile is
\begin{eqnarray}
S_m(\omega') &=& \langle\psi_m\vert \hat{s}_m^{\dagger}(\omega')\hat{s}_m(\omega') \vert\psi_m\rangle\label{eq:barium-frequency-profile}
\\
&=& \frac{2\tau_m}{\pi}\left[\frac{L_m}{1+ 4\tau_m^2 (\omega')^2}+\frac{1-L_m}{1+4 \tau_m^2 (\omega'-\omega_{hfs})^2}\right]\notag,
\end{eqnarray}
where $\vert \psi_m\rangle=\left(\hat{a}^{\dagger}_m+\hat{b}^{\dagger}_m\right)\vert 0\rangle$. Using a filter resonant at $\omega_m$, the filtered frequency mode profile becomes
\begin{equation}
S_{m,f}(\omega') =T(\omega') S_m(\omega'),
\label{eq:barium-frequency-filtered}
\end{equation}
where $T(\omega')=\langle f^*(\omega')f(\omega')\rangle$ \cite{olindo:filter} is the frequency-dependent transmission of light through the filter and
\begin{equation}
f(\omega') = \frac{\pi\kappa_m}{\pi\kappa_m-2i(\omega')}
\label{eq:filter}
\end{equation}
is the filter mode function \footnote{This is only approximately the filter function for the cavity in Ref.~\cite{olindo:filter}, since we only consider a single mode here because we assume the frequency spacing of the modes of the filter is much larger than that of the atom.} where $\kappa_m = \frac{\pi c}{\mathcal{F}l}$ is the cavity decay rate of the spectral filter with speed of light $c$, cavity finesse $\mathcal{F}$, and cavity length $l$ \footnote{For example, a $\kappa_m$ value equal to 50 MHz can be established using a cavity with $\mathcal{F}=188$ and $l=10$ cm.}. Both the unfiltered and filtered frequency mode profiles are shown in Fig.~\ref{fig:frequency-filtering} to exemplify the full effects of the filtering when including the relative strengths of the wanted and unwanted transitions.

With filtering effects accounted for in the frequency mode, we transition back into the temporal mode for further analysis. After filtering, a single emitted photon in spatial mode $m$ from $^{137}$Ba$^{+}$ may be described in the temporal mode by the electric field operator
\begin{widetext}
\begin{eqnarray}
\hat{E}^{+}_{m,f}(t) &=& \frac{1}{\sqrt{2\pi}} \int_{-\infty}^{\infty} f(\omega-\omega_m)\hat{s}_m(\omega-\omega_m) e^{-i\omega t}\,d\omega\label{eq:barium-elec-operator}
\\
										&=& \pi\kappa_m\sqrt{\tau_m}e^{-i\omega_m t}\Theta(t)\left[\frac{\sqrt{L_m}\left( e^{-\frac{t\pi\kappa_m}{2}} - e^{-\frac{t}{2\tau_m}}\right)}{1 - \pi\kappa_m\tau_m}\hat{a}_m + \frac{\sqrt{1 - L_m}\left(e^{-\frac{t}{2\tau_m} - i\omega_{hfs}t} - e^{-\frac{t\pi\kappa_m}{2}}\right)}{\pi\kappa_m\tau_m-2i\omega_{hfs}-1}\hat{b}_m\right]\notag.
\end{eqnarray}
\end{widetext}
In the likely situation that the photon mode profiles are nonidentical due to distinguishable filter cavities in each mode, quality of two-photon interference may benefit by imposing an intentional temporal offset in the arrival time of emitted photons at the BS. We thus incorporate a temporal mode-mismatch, $\Delta t$, in the electric field operators for emitted photons in mode $m=1$, so that $\hat{E}^{\pm}_{1,f}(t)\Rightarrow\hat{E}^{\pm}_{1,f}(t-\Delta t)$. Experimentally, this temporal offset may be manipulated by simply changing the pathlength of mode $m=1$. Utilizing the steps delineated in Sec.~\ref{sec:photon-interference-analysis}, relevant temporal mode profiles and JDP cases may now be derived using $\hat{E}^{\pm}_{1,f}(t-\Delta t)$ and $\hat{E}^{\pm}_{2,f}(t)$.

Further substituting the electric field operators of this subsection into the general analysis, we may determine the entanglement fidelity of this ion-ion system. Fidelity is compared to a net efficiency
\begin{equation}
P_{ions}(W) = \vartheta\left[\eta_{ions}(W)\left(\theta\xi D_{PMT}T_{fib}T_{opt}\right)^2\right],
\label{eq:net_efficiency}
\end{equation}
where $\vartheta = 0.25$ is the probability of detecting $\vert{\psi^{-}_{h\nu}}\rangle$, $\eta_{ions}(W)$ is the two-photon detection efficiency for this system defined by Eq.~\ref{eq:detection}, $\xi \approx 0.15$ is the relative branching ratio of $\sigma$-polarized emissions from $^{137}$Ba$^{+}$ transitions into the $F=1$ and $F=2$ hyperfine levels depicted in Fig.~\ref{fig:total-structure}b \cite{sherman-thesis,metcalf1999}, $\theta = 0.12$ is the collection efficiency of $^{137}$Ba$^{+}$ \cite{streed:angle_collection}, $D_{PMT} = 0.25$ is the quantum efficiency of each PMT, $T_{fib} = 0.30$ accounts for coupling and transmission efficiencies of necessary optical fibers, and $T_{opt} = 0.95$ is the transmission efficiency of necessary optics. The rate of entanglement generation is given by
\begin{equation}
\Gamma_{ions}(W)=\Gamma_{rep} P_{ions}(W),
\label{eq:success-rate}
\end{equation}
where we take $\Gamma_{rep}=5$ MHz based on recently demonstrated methods of qubit state preparation \cite{campbell-reprate}.
\begin{figure}
	\centering
	\includegraphics[width=1.0\columnwidth,keepaspectratio]{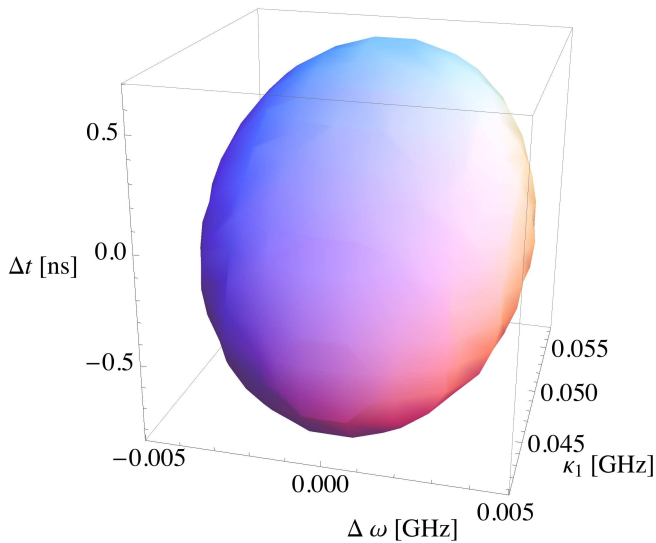}
	\caption{Region representing combinations of $\Delta\omega$, $\kappa_1$, and $\Delta t$ that permit $F(W=45 $ ns$,\Delta\omega)>99$\% and $\Gamma_{ions}(W=45$ ns$)>0.1$ Hz for the entanglement of two $^{137}$Ba$^{+}$ atoms. Here $\tau_1=\tau_2=8$ ns is the natural excited state lifetime and $\kappa_2=0.05$ GHz. It is clear that high fidelities and rates of entanglement can be achieved as long as $\vert \Delta\omega \vert < 0.005$ GHz, $0.040<\kappa_1<0.057$ GHz, and $-0.8 < \Delta t < 0.7$ ns.}
	\label{fig:ion-fidelity}
\end{figure}

Figure~\ref{fig:ion-fidelity} shows a solid region composed of coordinates $(\Delta\omega,\kappa_1,\Delta t)$, which provide fidelities $> 99$\% and rates of entanglement $>0.1$ Hz when detecting during a reasonable window $W=45$ ns. The remote entanglement therefore may maintain acceptable results even under unavoidable mode-mismatches caused by distinguishable filter cavities, especially when introducing experimentally optimal temporal offsets. In addition, the use of $\sigma$-polarized photons may enable greater photon collection efficiencies, and thus greatly increases entanglement rates, in the near future \cite{lindlein2007,luo:scalable_network}.

Although we have specifically analyzed the $^{137}$Ba$^{+}$ atom, the above procedure is generally applicable to any atomic ion with a low-lying $D$-state where transition wavelengths may be more practical for longer-distance transmission (e.g., Sr$^+$ or Ca$^+$).

\subsection{Two Solid-State Qubits}
\begin{figure*}
	\centering
	\includegraphics[width=2.0\columnwidth,keepaspectratio]{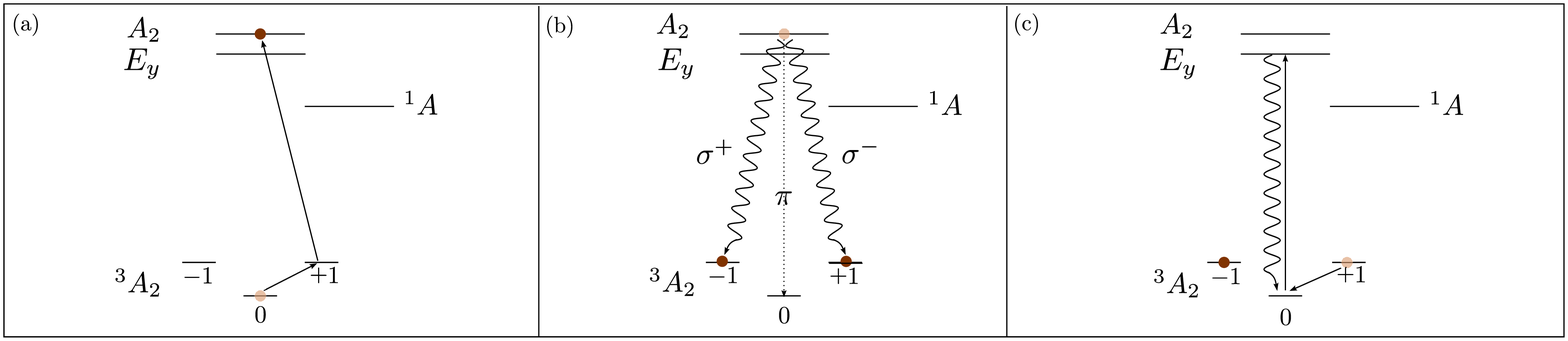}
	\caption{(a) After optical pumping to the $^{3}A_{2}$ $\vert m_s=0 \rangle$ state, a microwave pulse transfers population to $^{3}A_{2}$ $\vert m_s= +1 \rangle$ \cite{togan2010-nvc-photon}. Then, a 637 nm pulse excites the NV center into the $A_{2}$ excited state with natural lifetime $\tau_{0}=12$ ns. (b) The transition to $^{3}A_{2}$ $\vert m_s=\pm1 \rangle$ can establish spin-polarization entanglement between a spontaneously emitted photon and an NV center \cite{togan2010-nvc-photon}.  With cavity-coupling, only photons at the zero-phonon line with the wavelength of 637 nm will be observed. (c) Read-out of the NV center qubit spin state \cite{togan2010-nvc-photon}. Microwave radiation transfers population from $^{3}A_{2}$ $\vert m_s= +1 \rangle$ to $^{3}A_{2}$ $\vert m_s=0 \rangle$. Resonant light at 637 nm between $^{3}A_{2}\vert m_s=0 \rangle$ and $E_y$ causes fluorescence only if the original state was $^{3}A_{2}$ $\vert m_s= +1 \rangle$. Ideally, no scattered photons would be detected if the qubit was initially in the state $^{3}A_{2}$ $\vert m_s= -1 \rangle$.}
	\label{fig:nvc-entanglement}
\end{figure*}

In another extension of the generic protocol in Sec.~\ref{sec:initial-protocol} we consider the remote entanglement of two low-temperature solid-state matter qubits. Specifically, we apply our analysis to the entanglement of two nitrogen-vacancy (NV) centers through the interference of emitted photons at the zero-phonon line (ZPL) near 637 nm. Interference has recently been demonstrated between photons emitted by NV centers \cite{hanson:nv_interference,newlukin_nvc2011} without cavity-coupling. Due to a low emission probability at the ZPL \cite{togan2010-nvc-photon}, we instead consider the fabrication of a cavity resonant at the ZPL of each NV center for enhanced spontaneous emission and collection. Given unavoidable distinguishability between fabricated cavities, our following analysis demonstrates the limits of entangling NV centers in terms of the quality factor of each cavity. Following the recent experimental demonstration of entanglement between an NV center and a single photon \cite{togan2010-nvc-photon}, in Fig.\ref{fig:nvc-entanglement} we review the processes of excitation, emission, and detection for achieving entanglement following initialization by optical pumping into the $^{3}A_{2}$ $\vert m_s=0 \rangle$ state.

Cavity-coupling of an NV center may be necessary to enhance the emission rate of photons characterized by the ZPL wavelength $\lambda_{ZPL} = \lambda_m = 637$ nm. In order for this to be done with high efficiency, in this analysis we incorporate the fabrication of a photonic crystal cavity to the NV center corresponding to each mode $m$, which has been successfully demonstrated with a single NV center in previous experiments \cite{englund2010-cavity-nvc}. The general emission enhancement of each NV center caused by the cavity-coupling may be determined using the Purcell factor
\begin{equation}
\label{eq:purcell}
p(Q_m) = \frac{3 Q_m}{4\pi^2 V_{cav}}\left(\frac{\lambda_m}{n_{cav}}\right)^3,
\end{equation}
assuming no spatial or resonance offsets exist between the cavity and the NV center, where $Q_m$ is the quality factor, $\lambda_m = \lambda_{ZPL}$ is the resonant wavelength, $n$ is the index of refraction, and $V_{cav}$ is the effective cavity mode volume. Following the assumption that a photonic crystal cavity may be fabricated with $V_{cav} \sim \left(\frac{\lambda_m}{n_{cav}}\right)^3$ \cite{englund2010-cavity-nvc}, the relevant Purcell factor is
\begin{equation}
\label{eq:purcell-simplified}
p(Q_m) = \frac{3 Q_m}{4\pi^2}.
\end{equation}
The Purcell factor in Eq.~\ref{eq:purcell-simplified} may be utilized to compute the cavity-enhanced lifetime \cite{faraon:nvc_enhancement,fu:branching_ratio}
\begin{equation}
\label{eq:enhanced-nvc-lifetime}
\tau_m = \frac{\tau_{0}}{1+p(Q_m) \xi_{0,ZPL}},
\end{equation}
where $\xi_{0,ZPL} = 0.03$ \cite{faraon:nvc_enhancement,togan2010-nvc-photon} is the natural branching ratio of photon emissions at the ZPL.

Unlike the scenario in Sec.~\ref{sec:barium}, a cavity-coupled NV center may be described in the temporal mode with an electric field operator that is analogous to that expressed in Eq.~\ref{eq:general-operator}. However, a difference occurs due to the substitution of the righthand side of Eq.~\ref{eq:enhanced-nvc-lifetime} in the place of $\tau_m$, where distinguishability arises between each fabricated cavity-coupled NV center from the variability of $Q_m$.
\begin{figure}
	\centering
	\includegraphics[width=1.0\columnwidth,keepaspectratio]{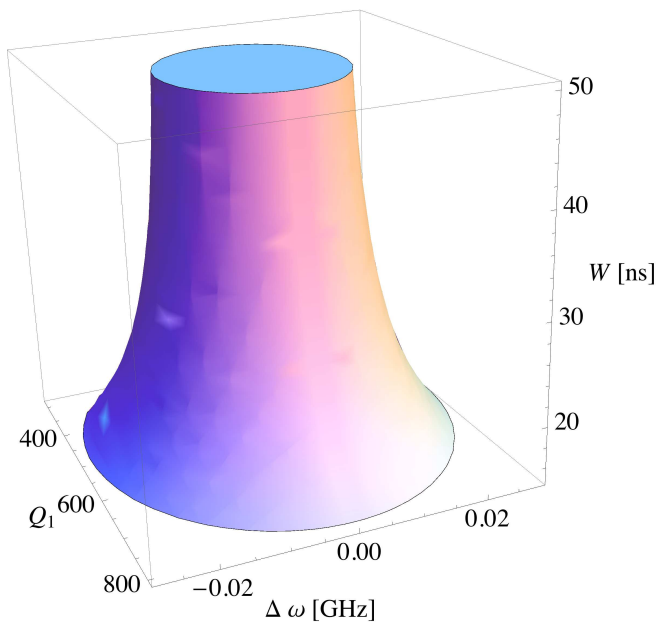}
	\caption{Region representing combinations of $Q_2$, $\Delta\omega$, and $W$ that permit $F(W,\Delta\omega)>99$\% and $\Gamma_{NVs}(W)>25$ Hz for the entanglement of two NV centers. Here we assume that $\tau_0=12$ ns is the natural excited state lifetime for both NV centers, and $Q_2=500$ characterizes the lower quality cavity. The results for this system are most similar to those in Sec.~\ref{sec:photon-interference-analysis} because we observe single-exponential decay profiles in both photons. Thus, it is clear that high fidelities and rates of entanglement can be reached for $\vert \Delta\omega \vert <0.025$ GHz, $400<Q_1<800$, and $W>15$ ns.}
	\label{fig:nvc-fidelity}
\end{figure}
Then, employing the steps outlined in Sec.~\ref{sec:photon-interference-analysis}, the necessary temporal mode profiles and JDPs may be derived for the calculation of entanglement fidelity and efficiency of entanglement. Fidelity of the entanglement of two cavity-coupled NV centers is compared to a net efficiency
\begin{equation}
P_{NVs}(W) = \vartheta\left[\eta_{NVs} (W)\xi_{p,1} \xi_{p,2} \beta_1\beta_2 \left(D_{PMT}T_{fib}T_{opt}\right)^2\right],
\label{eq:net_efficiency-2}
\end{equation}
where $\vartheta$, $T_{fib}$, $D_{PMT}$, and $T_{opt}$ are taken to be the same as in Sec.~\ref{sec:barium}, $\eta_{NVs}(W)$ is the two-photon detection efficiency of this system based on Eq.~\ref{eq:detection}, $\xi_{p,m} = \frac{\xi_{0,ZPL}\left(3Q_m+4\pi^2\right)\tau_{0}}{3Q_m\xi_{0,ZPL}+4\pi^2}$ is the cavity-enhanced branching ratio of the NV center emissions at the ZPL \cite{faraon:nvc_enhancement}, and $\beta_m = \frac{p(Q_m)}{p(Q_m)+1}$ is the fraction of photons emitted into the cavity mode in the weak-coupling regime, which we take as the effective collection efficiency \cite{barnes:collection_efficiency,gerard:collection_efficiency}. The entanglement rate for two cavity-coupled NV centers is then
\begin{equation}
\Gamma_{NVs}(W)=\Gamma_{rep} P_{NVs}(W),
\label{eq:success-rate-2}
\end{equation}
which follows the convention used in Eq.~\ref{eq:success-rate}, where here we consider $\Gamma_{rep} =100$ kHz \cite{togan2010-nvc-photon}. In Fig.~\ref{fig:nvc-fidelity} we illustrate a solid region composed of coordinates $(\Delta\omega,Q_1,W)$ with $Q_2=500$, where the protocol achieves fidelities $>99$\% and rates of entanglement $>25$ Hz. Thus, the remote entanglement of solid-state color centers has the ability to uphold exceptional results under distinguishability in cavity fabrication.

\subsection{Hybrid Qubits}
\label{sec:hybrid}
For our final extension to the general protocol in Sec.~\ref{sec:initial-protocol} we consider the hybrid entanglement of a low-temperature solid-state matter qubit and an atomic matter qubit. As aforementioned in Sec.~\ref{sec:introduction}, a proposal exists to establish hybrid entanglement between a semiconductor InAs quantum dot and a trapped $^{171}$Yb$^+$ atom using a single-photon interference scheme \cite{waks-hybrid}. In the following analysis we evaluate the hybrid entanglement between the same two quantum memories using the two-photon interference scheme presented in the above sections.
\begin{figure*}
	\centering
	\includegraphics[width=2.0\columnwidth,keepaspectratio]{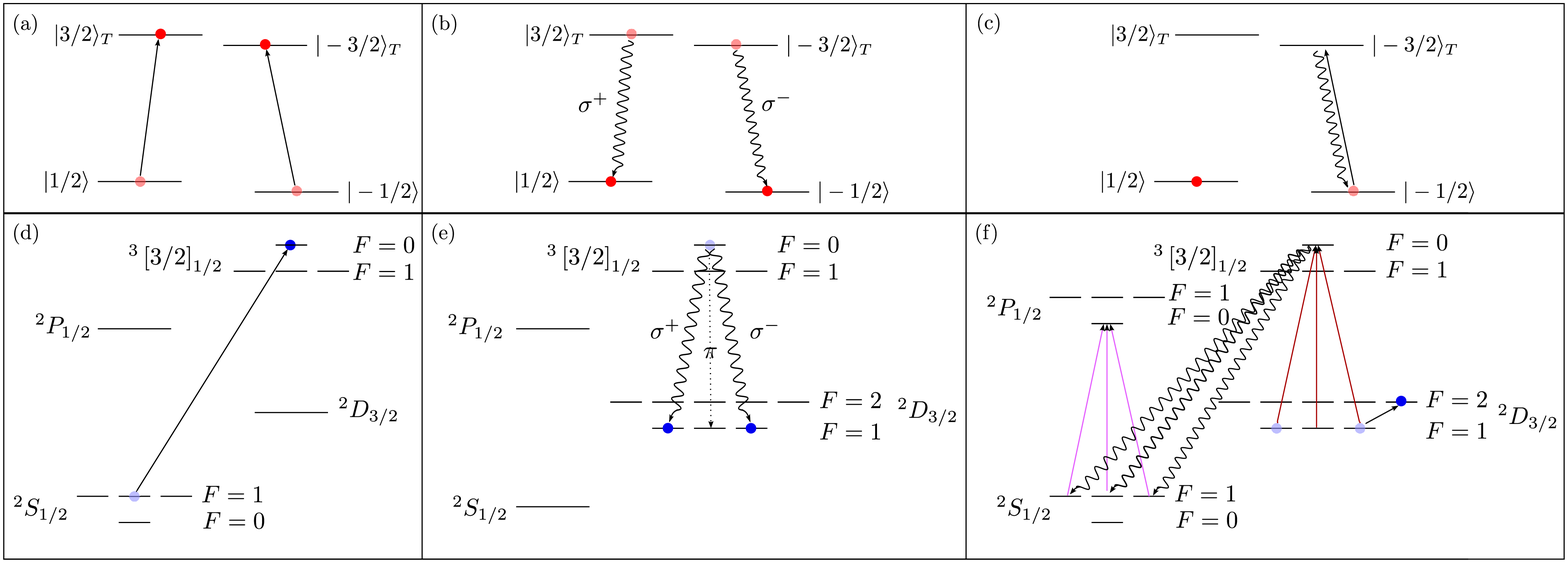}
	\caption{(a) After initializing the ground state of the quantum dot in the superposition $\frac{\vert 1/2\rangle+\vert -1/2\rangle}{\sqrt{2}}$, a laser pulse resonant at 935 nm excites the quantum dot into a superposition of trion spin states $\frac{\vert 3/2\rangle_T+\vert -3/2\rangle_T}{\sqrt{2}}$. (b) The superposition of trion states will decay after excited state lifetime $\tau_{QD}=0.3$ ns, emitting a single photon in a superposition of circular polarization states. (c) Read-out of the quantum dot qubit. A $\sigma^-$-polarized beam resonant at the $\vert -1/2\rangle$ to $\vert -3/2\rangle_T$ transition (935 nm) is applied to the quantum dot. If projected into the $\vert -1/2\rangle$ state, fluorescence will be detected, while if in the $\vert 1/2\rangle$ state, no fluorescence will be detected. (d) After initializing Yb$^+$ into $^2S_{1/2}\vert F=1,m_F=0\rangle$ using optical pumping, a laser pulse resonant at 297 nm excites the ion into $^3[3/2]_{1/2}\vert F=0,m_F=0\rangle$. (e) $^3[3/2]_{1/2}\vert F=0,m_F=0\rangle$ will decay with excited state lifetime $\tau_{Yb}=37.7$ ns into a superposition of spin states $^2D_{3/2}\vert F=1,m_F=\pm1\rangle$ while emitting a single photon in a superposition of circular polarization states; photons with $\pi$-polarization are excluded by observing along a quantization axis defined by an external magnetic field \cite{blinov-atom-photon}. (f) Read-out of the $^{171}$Yb$^+$ qubit. Microwave radiation transfers population from $^2D_{3/2}\vert F=1,m_F=+1\rangle$ to $^2D_{3/2}\vert F=1,m_F=+2\rangle$. Light resonant at the $^2D_{3/2}\vert F=1\rangle$ to $^3[3/2]_{1/2}\vert F=0\rangle$ transition (935 nm) and the $^2S_{1/2}\vert F=1\rangle$ to $^2P_{1/2}\vert F=0\rangle$ transition (370 nm) are applied to $^{171}$Yb$^+$. If the atom is projected into the $^2D_{3/2}\vert F=1,m_F=-1\rangle$, fluorescence will be detected, while if in $^2D_{3/2}\vert F=1,m_F=+1\rangle$, no fluorescence will be detected.}
	\label{fig:hybrid-entanglement}
\end{figure*}

Existing proposals for the remote entanglement of two charged quantum dots consider both single-photon interference \cite{childress:dots_singlephoton} and two-photon interference \cite{simon:dots_twophoton}. For two-photon interference, Ref.~\cite{simon:dots_twophoton} utilizes entanglement between emissions from trion decays and qubit spin states in the lowest levels. Our extended entanglement protocol includes a similar convention. We assume initialization of the quantum dot may be done by creating a superposition $\frac{\vert 1/2\rangle+\vert -1/2\rangle}{\sqrt{2}}$ between spin states using single-qubit rotation. The process of establishing matter-photon entanglement between an emitted photon and a quantum dot is described by Fig.~\ref{fig:hybrid-entanglement}(a)-(c). We assume an excited state lifetime of the trion level $\tau_{QD} \sim 0.3$ ns \cite{langbein:qd_lifetime,simon:dots_twophoton}; the transition wavelength of interest is 935 nm, which is near the relevant $^{171}$Yb$^+$ transition.

The protocol for matter-photon entanglement with $^{171}$Yb$^+$ employs the transition from $^3[3/2]_{1/2}$ to $^2D_{3/2}$ with emission wavelength 935 nm and $^3[3/2]_{1/2}$ excited state lifetime $\tau_{Yb} = 37.7$ ns \cite{berends:lifetime,olmschenk:quantum_logic}. The process for establishing matter-photon entanglement with $^{171}$Yb$^+$ is detailed in Fig.~\ref{fig:hybrid-entanglement}(d)-(f). 

As a result of the selection rules for both the selected quantum dot and $^{171}$Yb$^+$ transitions, spectral filtering is not required to use the $\sigma$-polarized transitions to the low-lying $D$ level as it was with $^{137}$Ba$^{+}$. However, we include a spectral filter only in the path of the quantum dot emissions to improve temporal mode-matching \footnote{Also, quantum dot emissions may exhibit inhomogeneous broadening. By utilizing a Fabry-Perot cavity with linewidth $< 1$ GHz we assume error due to inhomogeneous broadening negligible.}. With the effects of filtering accounted for, the quantum dot emits photons in mode $m=1$ that may be described with electric field operators $\hat{E}^{\pm}_{1,f}(t)$ based on Eq.~\ref{eq:barium-elec-operator} when $L_1=1$ and $^{171}$Yb$^+$ emits photons in mode $m=2$ that may be described with electric field operators $\hat{E}^{\pm}_{2}(t)$ based on Eq.~\ref{eq:general-operator}. Due to the different types of electric field operators for each qubit emission, we incorporate a temporal offset $\Delta t$ in the quantum dot emission so that its electric field operators become $\hat{E}^{\pm}_{1,f}(t)\Rightarrow\hat{E}^{\pm}_{1,f}(t-\Delta t)$, as is done in Sec.~\ref{sec:barium}.

As with the earlier selected systems, the derivations follow the step-by-step methods in Sec.~\ref{sec:photon-interference-analysis} to determine fidelity and two-photon detection efficiency.
\begin{figure}
	\centering
	\includegraphics[width=1.0\columnwidth,keepaspectratio]{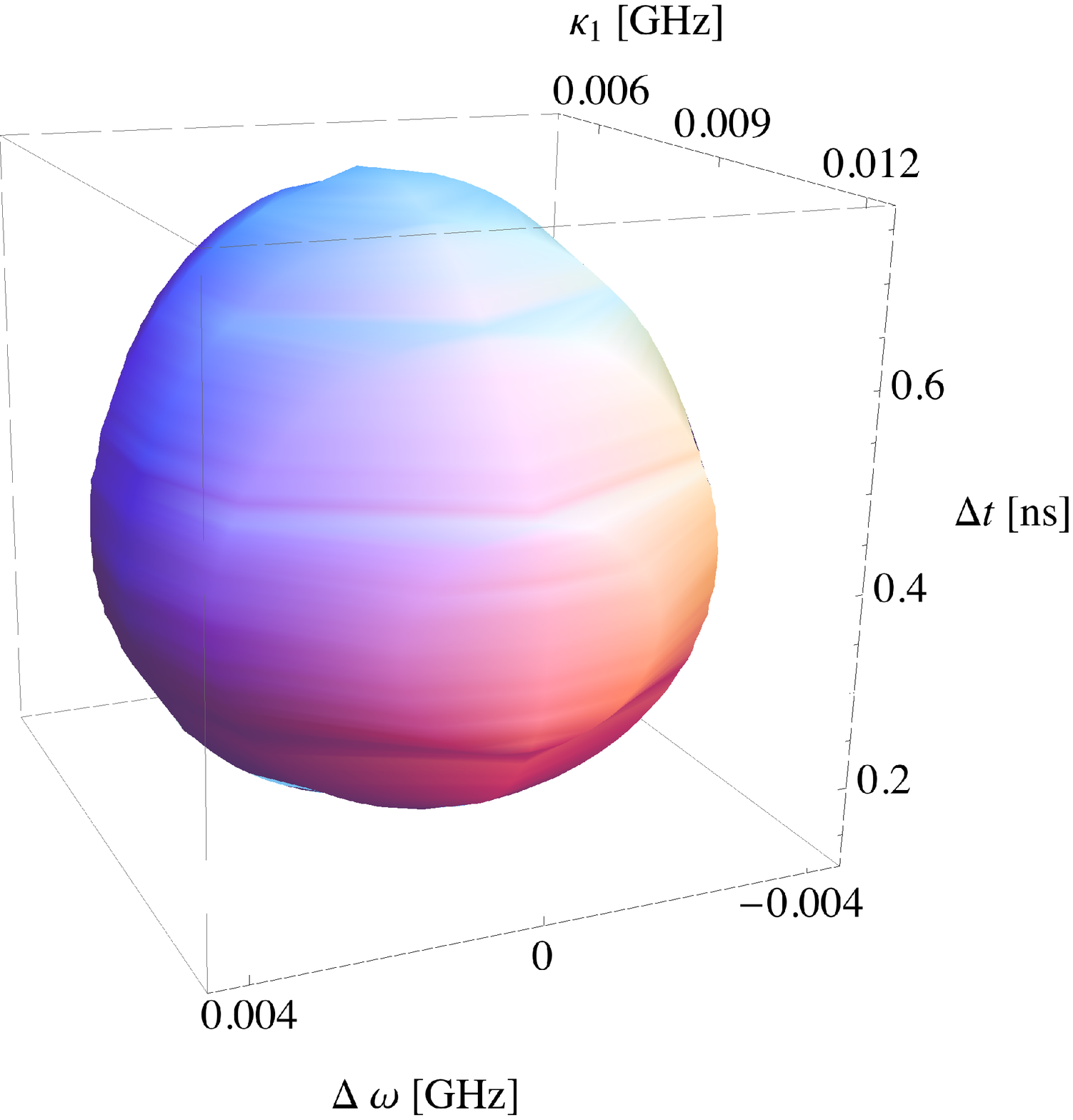}
	\caption{Region representing combinations of $\Delta\omega$, $\kappa_1$, and $\Delta t$ that permit $F(W=42$ ns$,\Delta\omega)>99$\% and $\Gamma_{hyb}(W=42$ ns$)>0.002$ Hz for the entanglement of a quantum dot and an $^{171}$Yb$^+$ atom. Here $\tau_1=0.3$ ns and $\tau_2=37.7$ ns are the natural excited state lifetimes of the quantum dot and $^{171}$Yb$^+$, respectively. Despite the vast differences in the emissions of the two qubits, high entanglement fidelities can still be achieved at reasonable rates of entanglement as long as $\vert \Delta\omega \vert < 0.004$ GHz, $0.005<\kappa_1<0.012$ GHz, and $0.1< \Delta t < 0.7$ ns.}
	\label{fig:hybrid-fidelity}
\end{figure}
The net efficiency is
\begin{equation}
P_{hyb}(W) = \vartheta\left[\eta_{hyb} (W)\xi_{1} \xi_{2} \theta_{1} \theta_{2} \left(D_{PMT}T_{fib}T_{opt}\right)^2\right],
\label{eq:net_efficiency-3}
\end{equation}
where $\vartheta$, $D_{PMT}$, $T_{fib}$, and $T_{opt}$ are taken to be the same as in Sec.~\ref{sec:barium}, $\eta_{hyb}(W)$ is the two-photon detection efficiency of this system based on Eq.~\ref{eq:detection}, $\xi_{1} \sim 1$ is the assumed branching ratio of the relevant quantum dot transition, $\xi_{2} = 0.018$ is the branching ratio of the relevant $^{171}$Yb$^+$ transition \cite{biemont:lifetime}, $\theta_1 = 0.10$ is the collection efficiency of quantum dot emissions \cite{davanco:collection_efficiency}, and $\theta_2=0.12$ is the collection efficiency of $^{171}$Yb$^+$ emissions \cite{streed:angle_collection}. The entanglement rate between a quantum dot and a single $^{171}$Yb$^+$ atom is then
\begin{equation}
\Gamma_{hyb}(W)=\Gamma_{rep} P_{hyb}(W),
\label{eq:success-rate-3}
\end{equation}
which follows the convention used in Eq.~\ref{eq:success-rate}, where we take $\Gamma_{rep} =5$ MHz \cite{campbell-reprate}. In Fig.~\ref{fig:hybrid-fidelity} we illustrate a solid region composed of coordinates $(\Delta\omega,\kappa_1,\Delta t)$, where fidelities $>99$\% and rates of entanglement $>0.002$ Hz are possible when detecting during a reasonable window $W=42$ ns. Thus, the remote entanglement of these hybrid qubits has the ability to uphold interesting results when appropriate experimental techniques are implemented.

While we have analyzed the entanglement specifically between a quantum dot and $^{171}$Yb$^+$, a similar protocol may be utilized for the entanglement between a quantum dot and either $^{87}$Sr$^+$ or $^{43}$Ca$^+$. Each of these ions behave like $^{137}$Ba$^+$ as in Sec.~\ref{sec:barium}, and thus frequency filtering may be required for two-photon interference. The relevant transitions of $^{87}$Sr$^+$ and $^{43}$Ca$^+$ have much broader linewidths than $^{171}$Yb$^+$, which may allow for improved pairing of fidelity and rate of entanglement \footnote{It should also be possible to implement the protocol with $^{88}$Sr$^+$ or $^{40}$Ca$^+$ by employing qubit shelving techniques previously demonstrated \cite{kaler:other_system}.  Here the atom-photon entanglement can be generated via the $^2P_{3/2}$ to $^2D_{5/2}$ decay, mapping one of the $^2D_{5/2}$ states back to $^2S_{1/2}$, and performing fluorescence detection on the $^2S_{1/2}$ to $^2P_{1/2}$ (and $^2P_{1/2}$ to $^2D_{3/2}$) transition.}. Additionally, cavity-coupling techniques \cite{keller:calcium,barros:singlephoton} could be used to substantially increase the fidelity and efficiency of the system, at the cost of increased experimental complexity. Finally, larger frequency differences between qubits may be mediated by the tunability of solid-state systems \cite{faraon_APL2007,bassett_PRL2011,patel_nature2010} or through translation of photon frequencies \cite{raymer_PRL2010}.

\section{Conclusion}

We have shown that a robust two-photon interference protocol can be used to entangle distinguishable quantum memories with high fidelities and rates.  A general analysis demonstrates that entanglement fidelities $>99$\% and detection efficiencies $>90$\% are possible even with significant mismatches in the frequency and temporal characteristics of the photons.  By applying this analysis to three relevant systems, we illustrate both the utility of the calculations and their applicability to current experimental efforts.  Moreover, the general analysis presented here can be easily extended to other pairs of quantum memories.  Overall, this evaluation elucidates the possibility of establishing remote entanglement between hybrid quantum memories that relies only on passive filtering and the passive stability inherent in the two-photon interference scheme.  Ultimately, this type of stalwart architecture may be crucial for the realization of long-distance quantum communication and distributed quantum computation.

\begin{acknowledgments}
We thank Dirk Englund and Dmitry Budker for helpful discussions concerning the cavity-coupling of NV centers and James V. Porto for useful comments on this manuscript. A.M.D. acknowledges the faculty at the Loudoun County Public Schools Academy of Science. S.O. acknowledges support from the National Research Council (NRC) Research Associateship program.
\end{acknowledgments}

%

\end{document}